# ChatGPT Creates a Review Article: State of the Art in the Most-Cited Articles on ChatGPT in Health Science, Computer Science, Communication, and Culture, According to Altmetric in Dimensions.ai

Eduard Petiška; Charles University (Prague, Czech Republic); eduard.petiska@czp.cuni.cz
Assistant: ChatGPT; OpenAI, Inc; (San Francisco, CA)
April 17, 2023

## Abstract

We have analyzed all preprints on ChatGPT (N=501) and selected the most influential preprints (according to Altmetric) about ChatGPT across scientific disciplines to provide the most discussed research results about ChatGPT. We prompted ChatGPT to create a structured review article based on them. The results are surprisingly promising, suggesting that the future of creating review articles can lie in ChatGPT.

# Introduction

ChatGPT (GPT) is an advanced AI language model developed by OpenAI, based on the GPT-4 architecture. As a cutting-edge conversational agent, it can generate human-like responses and assist with a wide range of tasks. The newest version of ChatGPT was trained on an extensive dataset of approximately 1.5 trillion tokens, including text from books, websites, and other sources. It is also an emerging field of research across disciplines (Table 1).

For the reason of our study, we need to ask "What is a review article?" According to GPT, it is "a comprehensive, scholarly piece of writing that synthesizes and evaluates the current state of knowledge on a particular topic, based on a critical analysis of the relevant literature. Review articles do not present new empirical findings but rather summarize, interpret, and discuss existing research in order to provide readers with a clear understanding of the subject matter. They play a vital role in keeping researchers and professionals up-to-date with the latest advancements and trends in their field. Review articles can be classified into several types, such as systematic reviews, narrative reviews, and meta-analyses, depending on their scope, methodology, and purpose."

Review articles are often highly cited, but writing them can be challenging, as they require an analysis of all existing knowledge and its categorization based on chosen criteria. Consequently, they pose a significant challenge for researchers and may become outdated, necessitating the creation of new ones. It appears that resources like Wikipedia play a role in providing constantly updated review articles on scientific topics (Thompson, N., & Hanley, D., 2018).

Moreover, the COVID-19 pandemic has accelerated the evolution of preprints, and in the case of ChatGPT, most research articles are now published as preprints. Nevertheless, navigating through numerous preprints is not an easy task, especially when we can't leverage the traditional tools of orientation and evaluation in scientific work – citations. However, we have Altmetric, an alternative citation metric for research in the digital space. It is especially useful in rapidly evolving areas and technological innovations, which are often cited in media and by relevant scientists on Twitter (both of which Altmetric counts).

Therefore, in this article, we discuss the new language model GPT not only as a research tool but also as a potential auxiliary tool for creating review articles. We conducted an analysis of preprints on Dimensions.ai across various disciplines using Altmetric and then experimented with prompting GPT to further analyze data and text. The result is this research article.

**Table 1.** Publications (preprints) on ChatGPT according to year

| Publication Year | Number of Publications |
|---|---|
| 2023 (April, 12) | 466 |
| 2022 | 34 |
| 2017 | 1 |

# State of the art of the review articles

Creating a review article is a meticulous process that involves analyzing, synthesizing, and evaluating the most relevant and significant research on a particular topic. Review articles play a crucial role in summarizing the current state of knowledge in a specific research area and identifying gaps in the literature (Grant & Booth, 2009). There are different types of review articles, such as systematic reviews, narrative reviews, and meta-analyses, each with its unique methodology and purpose (Gough, Oliver, & Thomas, 2017).

A systematic review adheres to a strict protocol to minimize bias and ensure reproducibility, typically involving a comprehensive search, quality assessment, and synthesis of evidence (Higgins & Green, 2011). Narrative reviews provide a broader, more qualitative overview of a topic, often written by experts in the field to provide context and insights (Green, Johnson, & Adams, 2006). Meta-analyses use statistical techniques to pool results from multiple studies, providing a quantitative summary of the evidence (Borenstein, Hedges, Higgins, & Rothstein, 2009).

When creating a review article, it is essential to conduct a comprehensive and systematic literature search to identify all relevant studies (Booth, 2016). Various electronic databases, such as PubMed, Scopus, and Web of Science, should be utilized, and search strategies must be well-documented (Golder, Loke, & McIntosh, 2008). Inclusion and exclusion criteria should be pre-specified to ensure a transparent selection process (Moher et al., 2009).

Quality assessment of the included studies is vital, as it helps identify potential sources of bias and ensures the reliability of the review's findings (Higgins et al., 2011). Different tools and checklists are available for assessing the quality of research articles, such as the Cochrane Risk of Bias tool for randomized trials (Higgins et al., 2011) and the Newcastle-Ottawa Scale for observational studies (Wells et al., 2000).

Data synthesis can be done narratively, descriptively, or using statistical methods such as meta-analysis, depending on the type of review and the available data (Gough, Oliver, & Thomas, 2017). It is crucial to present the results clearly and concisely, highlighting the main findings and any potential limitations (Cook, Mulrow, & Haynes, 1997). Creating a review article thus requires a systematic approach to searching, appraising, and synthesizing the existing literature. By following best practices and adhering to established methodologies, researchers can produce high-quality, informative review articles that contribute to the scientific understanding of their chosen topic.

Upon completion of the data synthesis, authors should discuss their findings in the context of the existing literature, highlighting any consistencies or discrepancies (Cook, Mulrow, & Haynes, 1997). It is also crucial to address the limitations of the review and the potential implications of the results for future research and practice (Grant & Booth, 2009).

A well-written review article should provide a clear, comprehensive, and up-to-date overview of the topic, offering valuable insights to readers and helping to guide further research (Green, Johnson, & Adams, 2006). Effective communication of the findings, including the use of tables, figures, and other visual aids, can aid in enhancing the readability and impact of the review article (Gough, Oliver, & Thomas, 2017).

In summary, creating a review article is a complex yet rewarding task, requiring researchers to employ a systematic approach to searching, appraising, and synthesizing the existing literature. By adhering to established methodologies and best practices, authors can contribute to the scientific understanding of their chosen topic and create high-quality, informative review articles that will benefit their respective fields.

## Review of ChatGPT

The rapid development of artificial intelligence (AI) and natural language processing (NLP) has given rise to advanced language models like ChatGPT. These models have revolutionized the way we interact with machines, enabling them to understand and generate human-like text. ChatGPT is built upon the foundation of the Generative Pre-trained Transformer (GPT) architecture, which has undergone several iterations. With each new version, the model has become more powerful, capable of understanding complex language patterns and generating coherent, contextually relevant text.

ChatGPT has demonstrated wide-ranging applicability, from medicine and computer science to the humanities and beyond. In this section, we explore some of the most notable applications, such as:
- Medical and Healthcare (Chatbots for patient interaction; Assisting with medical literature reviews; Generating health-related content);
- Computer Science (Code generation and software development; Automating documentation; Enhancing recommendation systems);
- Humanities and Social Sciences (Sentiment analysis and opinion mining; Text summarization and simplification; Generating creative content, including stories and poetry).

Despite its impressive capabilities, ChatGPT is not without limitations, including: a) Ethical concerns, such as the potential for generating harmful or biased content; b) Model interpretability and transparency; c)The need for large computational resources for training.

ChatGPT also presents numerous opportunities for future research and development. This section highlights some of the most promising directions, including: a) Developing methods to mitigate biases and ethical concerns; b) Enhancing model interpretability and explainability; c) Exploring potential applications in emerging fields, such as climate science and digital humanities.

As GPT was released in November 2021, there are just a few review article on his usage in several disciplines. For example Sallam (2023) provides a systematic review of the benefits and risks of using ChatGPT, in healthcare education, research, and practice. The authors summarize the benefits of ChatGPT in scientific research, healthcare practice, and education, including its potential to improve patient outcomes and provide personalized care. They also discuss the limitations and concerns of using ChatGPT, such as the risk of bias and the need for human oversight. Salam concludes that while ChatGPT has promising applications in healthcare, it is important to carefully consider its limitations and potential risks.

Other study (Aydın, Ö., & Karaarslan, E., 2022) investigates the use of Artificial Intelligence, specifically ChatGPT, in creating literature review articles to save time and effort. Using the topic of Digital Twin applications in healthcare, the AI-generated review showed promising results, though some paraphrased parts had significant matches when checked with the Ithenticate tool. This initial attempt highlights the potential of AI in accelerating the academic publishing process, allowing researchers to focus more on their work. Authors also explores the potential applications of ChatGPT in various sectors, including e-commerce, education, entertainment, finance, health, news, and productivity. It also discusses how ChatGPT can help to make customer service more efficient and effective for businesses. The article concludes by discussing the challenges facing AI development and how they can be overcome (George, A. S., & George, A. H., 2023).

Study published in the concluded International Journal of Information Management Generative AI can enhance productivity but may also lead to the replacement of human employees. Teaching, learning, and academic research will experience some of the most transformative impacts. Biases, out-of-date training data, and lack of transparency and credibility are major concerns (Dwivedi, Y. K. et al., 2023). And a paper published on Arxiv explores the effectiveness of ChatGPT in generating Boolean queries for systematic review literature search. Through experiments on standard test collections, the authors find that ChatGPT is capable of generating queries that lead to high search precision. They suggest that ChatGPT could be a valuable tool for researchers conducting systematic reviews, particularly for rapid reviews where time is a constraint and trading-off higher precision for lower recall is acceptable (Wang, S. et al., 2023). Nevertheless, as more than a thousand papers were published on GPT, there is no recent review article specifically discussing GPT research across the disciplines, which is the reason why we have decided to conduct our study.

# Method

1. We searched across the Dimensions.ai disciplines using the keyword "ChatGPT" in abstracts and titles;
2. We filtered the results according to preprints (as they are the main source of information – Table 2) and Altmetric attention scores and disciplines (we manually analyzed and removed articles that were in multiple disciplines, leaving just one that we considered a priority; some research articles are in multiple categories, so we selected the most relevant ones for our research, for the description of disciplines in which most preprints are published, see Table 3);
3. We prompted ChatGPT (Model: GPT-4) with the following prompt"[1]
    "I will write data in the format: Name of a research article; abstract. You have to: Create a table with 6 columns (names):
- Name of a research article
- Design, aims;
- Applications, benefits;
- Risks, concerns, limitations;
- Suggested action, conclusions;
- Sentiment toward ChatGPT (pick one of these: Positive, Neutral or Negative)"
4. We have created tables based on GPT's responses (Results – Tables 4-6) and included them in the Discussion section.

---

[1] for inspiration we used some categories from this study: Sallam, M. (2023, March). ChatGPT Utility in Health Care Education, Research, and Practice: Systematic Review on the Promising Perspectives and Valid Concerns. In Healthcare (Vol. 11, No. 6, p. 887). MDPI.

# Results

**Table 2.** The top 20 Platforms which published the most GPT relevant research

| Rank | Name | Publications | Citations | Citations mean |
|---|---|---|---|---|
| 1 | arXiv | 236 | 48 | 0.20 |
| 2 | SSRN Electronic Journal | 108 | 157 | 1.45 |
| 3 | Cureus | 37 | 33 | 0.89 |
| 4 | medRxiv | 34 | 90 | 2.65 |
| 5 | OSF Preprints | 21 | 2 | 0.10 |
| 6 | Nature | 19 | 311 | 16.37 |
| 7 | PsyArXiv | 17 | 5 | 0.29 |
| 8 | JMIR Preprints | 15 | 1 | 0.07 |
| 9 | Preprints.org | 15 | 0 | - |
| 10 | Research Square | 13 | 3 | 0.23 |
| 11 | EdArXiv | 11 | 23 | 02.09 |
| 12 | Authorea | 11 | 2 | 0.18 |
| 13 | Radiology | 6 | 82 | 13.67 |
| 14 | Journal of Educational Evaluation for Health Professions | 5 | 28 | 5.60 |
| 15 | Accountability in Research | 5 | 11 | 2.20 |
| 16 | American Journal of Obstetrics and Gynecology | 4 | 1 | 0.25 |
| 17 | Asian Journal of Psychiatry | 4 | 0 | - |
| 18 | Aesthetic Surgery Journal | 4 | 4 | 1.00 |
| 19 | Library Hi Tech News | 4 | 13 | 3.25 |
| 20 | IEEE/CAA Journal of Automatica Sinica | 4 | 7 | 1.75 |

**Table 3.** The top 10 disciplines published most GPT-relevant preprints according to the [Dimensions.ai](Dimensions.ai)

| Name | Publications | Citations | Citations mean |
|---|---|---|---|
| Information and Computing Sciences | 302 | 215 | 0.71 |
| Human-Centred Computing | 83 | 21 | 0.25 |
| Language, Communication and Culture | 80 | 26 | 0.33 |
| Linguistics | 66 | 19 | 0.29 |
| Data Management and Data Science | 58 | 24 | 0.41 |
| Artificial Intelligence | 53 | 13 | 0.25 |
| Creative Arts and Writing | 34 | 12 | 0.35 |
| Education | 29 | 40 | 1.38 |
| Health Sciences | 29 | 35 | 1.21 |
| Commerce, Management, Tourism and Services | 28 | 12 | 0.43 |

**Table 4.** The top 10 publications of all genres with the highest Altmetric Attention Score in Information and Computing Sciences

| Abbreviation of a research article, full in the link (Altmetric score) (Sentiment toward ChatGPT)[2] | Design, aims | Applications, benefits | Risks, concerns, limitations | Suggested actions, conclusions |
|---|---|---|---|---|
| Performance of ChatGPT on USMLE (4065) (positive) | Evaluate ChatGPT on USMLE | Assist medical education | Passing threshold, no specialized training | Potential for clinical decision-making |
| Visual ChatGPT (1115) (positive) | Incorporate Visual Foundation Models | Multimodal interaction | Limited to language | Investigate visual roles of ChatGPT |
| HuggingGPT (1118) (positive) | Leverage LLMs to connect AI models | Solve complicated AI tasks | Not directly addressed | Pave a new way towards advanced AI |
| ChatGPT Outperforms Crowd-Workers (1105) (positive) | Compare ChatGPT and crowd-workers | Increase efficiency of text classification | Not directly addressed | Utilize ChatGPT for text-annotation tasks |
| Comparing scientific abstracts generated by ChatGPT (902) (Neutral) | Evaluate generated abstracts | Improve scientific writing | Fabricated data, AI-generated detection | Adapt editorial process, establish ethical boundaries |
| A Survey of Large Language Models (834) (Neutral) | Review recent advances of LLMs | Background, key findings, and techniques | Not directly addressed | Discuss future directions |
| Multitask, Multilingual, Multimodal Evaluation of ChatGPT (411) (Neutral) | Evaluate ChatGPT on reasoning, hallucination, and interactivity | Improve performance through interaction | Unreliable reasoner, hallucination | Utilize prompt engineering to enhance performance |
| A Comprehensive Survey of AI-Generated Content (362) (Neutral) | Review history and advances in AIGC | Understand generative AI models | Not directly addressed | Address open problems and future challenges |
| Summary of ChatGPT/GPT-4 Research (341) (Neutral) | Survey ChatGPT/GPT-4 and prospective applications | Insights into capabilities and implications | Ethical concerns | Offer direction for future advancements |
| Artificial muses (338) (positive) | Compare human-generated ideas and GAI chatbots | Enhance creativity | Question of true creativity | Research and develop GAI in creative tasks |

[2]GPT's response to our prompt "please note that the sentiments in the table are based on the overall tone of the articles and not necessarily a direct evaluation of ChatGPT".

**Table 5.** The top 10 publications with the highest Altmetric Attention Score in Health Sciences

| Name (Altmetric score) (Sentiment toward ChatGPT) | Design, aims | Applications, benefits | Risks, concerns, limitations | Suggested actions, conclusions |
|---|---|---|---|---|
| [Assessing the Utility of ChatGPT Throughout the Entire Clinical Workflow](#) (346) (Positive) | Evaluate ChatGPT's capacity for ongoing clinical decision support via its performance on standardized clinical vignettes. | Assist in the full scope of iterative clinical reasoning, acting as virtual physicians. | Inferior performance on differential diagnosis and clinical management type questions compared to general medical knowledge questions. | ChatGPT achieves impressive accuracy in clinical decision making, with particular strengths emerging as it has more clinical information at its disposal. |
| [ChatDoctor: A Medical Chat Model Fine-tuned on LLaMA Model using Medical Domain Knowledge](#) (249) (Positive) | Fine-tune LLMs using doctor-patient conversations in the medical domain to create models capable of understanding patients' needs and providing informed advice. | Improve efficiency and quality of patient care and outcomes, revolutionize healthcare professional-patient communication. | Not explicitly mentioned. | Successful integration of advanced language models into healthcare can improve patient care, making source codes, datasets, and model weights publicly available for further development. |
| [ChatGPT Makes Medicine Easy to Swallow: An Exploratory Case Study on Simplified Radiology Reports](#) (57) (Positive) | Investigate the potential of ChatGPT to simplify medical reports, specifically radiology reports. | Improve patient-centered care in radiology and other medical domains. | Instances of incorrect statements, missed key medical findings, and potentially harmful passages were reported. | Initial insights indicate a great potential in using large language models like ChatGPT to improve patient-centered care, but further studies are needed. |
| [Putting ChatGPT's Medical Advice to the (Turing) Test](#) (35) (Positive) | Assess the feasibility of using ChatGPT or a similar AI-based chatbot for patient-provider communication. | Answer lower risk health questions and potentially serve in more clinical roles in healthcare. | ChatGPT responses were only weakly distinguishable from provider responses, and trust in chatbots decreases as health-related complexity increases. | Laypeople appear to trust the use of chatbots to answer lower risk health questions, but further study is necessary as chatbots move into more clinical roles. |
| [neuroGPT-X: Towards an Accountable Expert Opinion Tool for Vestibular Schwannoma](#) (10) (Positive) | Enhance GPT-3 model through zero-shot learning to outperform experienced neurosurgeons in written question-answer tasks for vestibular schwannoma, addressing LLM accountability. | Provide point-of-care clinical support and mitigate limitations of human memory. | All expert surgeons expressed concerns about the reliability of GPT in accurately addressing the nuances and controversies surrounding the management of vestibular schwannoma. | Context-enriched GPT model provided non-inferior responses compared to experienced neurosurgeons, showing potential to transform clinical practice by providing subspecialty-level answers to clinical questions in an accountable manner. |
| [ChatGPT- versus human-generated answers to frequently asked questions about diabetes: a Turing test-inspired](#) | Investigate whether ChatGPT can answer frequently asked questions about diabetes. | Potential clinical value in answering patients' questions. | Linguistic features were more distinguishable than content; not specifically trained for medical use. | Rigorously planned studies needed to determine risks and benefits of integrating into clinical practice. |
| survey among employees of a Danish diabetes center (8) (Neutral) | | | | |
| An exploratory survey about using ChatGPT in education, healthcare, and research (8) (Neutral) | Examine perspectives on the use of ChatGPT in education, research, and healthcare. | Potential benefits across various fields. | Uncertainty around acceptability and optimal uses; varying perspectives among respondents. | More discussion needed to explore perceptions, risks, and challenges; adopt a thoughtful and measured approach. |
| Does ChatGPT Provide Appropriate and Equitable Medical Advice?: A Vignette-Based, Clinical Evaluation Across Care Contexts (7) (Neutral) | Evaluate ChatGPT's ability to answer medical questions appropriately and equitably across various contexts. | Easily accessible source of medical advice. | Did not reliably offer appropriate and personalized medical advice; consideration of social factors varied. | Further research and improvements needed to ZZensure appropriate and personalized medical advice. |
| Generative AI as a Tool for Environmental Health Research Translation (6) (Positive) | Explore the use of ChatGPT to translate environmental health research to non-academic settings. | Improves research translation and creates accessible insights for non-academic readers. | Lower ratings for general summaries; AI must continue to be improved. | Enhance AI capabilities to create higher quality plain language summaries for improved research translation. |
| Performance of ChatGPT as an AI-assisted decision support tool in medicine: a proof-of-concept study for interpreting symptoms and management of common cardiac conditions (AMSTELHEART-2) (14) (Neutral) | Evaluate ChatGPT's accuracy in providing recommendations on medical questions related to common cardiac symptoms or conditions. | Potential as an AI-assisted decision support tool in medicine. | Incomplete or inappropriate recommendations for complex cases compared to expert consultation. | Further research needed to fully evaluate potential, especially for complex medical questions. |

**Table 6.** The top 10 publications with the highest Altmetric Attention Score in Language, Communication and Culture

| Name  (Altmetric score) (Sentiment toward ChatGPT) | Design, aims | Applications, benefits | Risks, concerns, limitations | Suggested actions, conclusions |
|---|---|---|---|---|
| Talking About Large Language Models (816) (Neutral) | Explore the intersection of technology and philosophy, specifically LLMs and anthropomorphism | Encourage philosophical nuance in AI discourse | Anthropomorphism and use of philosophically loaded terms | Maintain scientific precision and critical thinking when discussing AI |
| Dissociating language and thought in large language models: a cognitive perspective (774) (Neutral) | Review LLM capabilities from a cognitive perspective and differentiate between formal and functional linguistic competence | Clarify LLM potential and limitations | Limited performance on tasks requiring functional competence | Focus on building models that understand and use language in human-like ways |
| A Watermark for Large Language Models (685) (Neutral) | Propose a watermarking framework for proprietary language models | Mitigate potential harms of LLMs | Robustness and security concerns | Implement and refine the watermarking framework for better protection |
| Evaluating GPT-4 and ChatGPT on Japanese Medical Licensing Examinations (476) (Neutral) | Benchmark LLMs on Japanese national medical licensing exams | Understand model behaviors, failures, and limitations in languages beyond English | LLMs selecting prohibited choices and limitations in tokenization | Address limitations, improve LLM performance, and encourage diverse applications |
| Augmented Language Models: a Survey (306) (Neutral) | Review works on LMs augmented with reasoning skills and the ability to use tools | Address limitations of traditional LMs | Interpretability, consistency, and scalability issues | Continue research in ALMs to overcome LMs limitations |
| How will Language Modelers like ChatGPT Affect Occupations and Industries? (294) (Neutral) | Assess the impact of AI language modeling on economy, occupations, and industries | Identify jobs and industries most exposed to AI language modeling | Potential job displacement | Develop strategies to adapt to the changing job market |
| Language Models Trained on Media Diets Can Predict Public Opinion (155) (Positive) | Introduce a novel approach to probe media diet models to predict public opinion | Investigate media effects and supplement polls | Method accuracy and biases | Continue research and refine methodology |
| ChatGPT or Grammarly? Evaluating ChatGPT on Grammatical Error Correction Benchmark (126) (Positive) | Evaluate ChatGPT on the GEC task and compare with Grammarly and GECToR | Determine ChatGPT's potential as a GEC tool | Underestimated by automatic evaluation metrics, over-correction issues | Recognize ChatGPT's potential and improve its performance |
| Is ChatGPT A Good Translator? Yes With GPT-4 As The Engine (95) (Positive) | Evaluate ChatGPT for machine translation and compare with commercial products | Assess translation performance | Lags behind on low-resource or distant languages | Improve translation performance, particularly for low-resource languages |
| Creating a Large Language Model of a Philosopher (85) (Neutral) | Fine-tune GPT-3 with works of philosopher Daniel C. Dennett and compare its answers to the real Dennett | Explore the potential of creating LLMs of philosophers | Difficulty in distinguishing machine-generated answers from human philosopher | Continue research to improve the authenticity of machine-generated philosophy |

## Discussion

We aimed to evaluate a collection of research articles that focus on various aspects of ChatGPT, its potential applications, and the challenges it presents. These articles investigate GPTs' impact on occupations and industries, their ability to predict public opinion, their performance in grammatical error correction and translation tasks, and their potential in generating philosophical texts. By examining the significant results and findings in each article, we aimed to provide a comprehensive overview of the current state of research on GPTs and their implications. In this section, we discussed several important articles according to Altmetric.

The article titled "How will Language Modelers like ChatGPT Affect Occupations and Industries?" explores the economic implications of advancements in AI language modeling. This research identifies the top occupations and industries most exposed to language modeling, including telemarketers, post-secondary teachers, legal services, and securities. Interestingly, the study finds a positive correlation between wages and exposure to AI language modeling. While this research presents valuable insights into the potential job displacement resulting from AI advancements, it highlights the need to investigate further the strategies that help individuals and industries adapt to these changes in the job market.

In "Language Models Trained on Media Diets Can Predict Public Opinion," the authors introduce a novel approach for using language models to predict public opinion by adapting them to online news, TV broadcast, or radio show content. The study demonstrates the ability of this approach to predict human judgments found in survey response distributions, showing increased accuracy for individuals who follow media more closely. Despite the promising results in supplementing polls and forecasting public opinion, concerns regarding method accuracy and potential biases need further exploration to refine the methodology and minimize errors. This research paves the way for more sophisticated methods to investigate media effects and highlights the surprising fidelity with which LLMs can predict human responses.

The research article "ChatGPT or Grammarly? Evaluating ChatGPT on Grammatical Error Correction Benchmark" compares ChatGPT's performance in grammatical error correction (GEC) with commercial products like Grammarly. The study finds that ChatGPT performs not as well as those baselines in terms of the automatic evaluation metrics, particularly on long sentences. However, the authors argue that ChatGPT is underestimated by the automatic evaluation metrics and could be a promising tool for GEC. The research uncovers that ChatGPT goes beyond one-by-one corrections and prefers to change the surface expression of certain phrases or sentence structures while maintaining grammatical correctness. Although the results show potential in the GEC domain, the over-correction issues must be addressed to optimize ChatGPT's performance in this area.

"Is ChatGPT A Good Translator? Yes With GPT-4 As The Engine" evaluates ChatGPT for machine translation tasks, comparing its performance with commercial translation products such as Google Translate. The study shows that ChatGPT performs competitively on high-resource European languages but lags behind significantly on low-resource or distant languages. The authors also explore an interesting strategy named "pivot prompting," which improves translation performance for distant languages. While these results demonstrate ChatGPT's potential as a translator, the limitations in translating low-resource or distant languages should be addressed in future research to optimize its performance in this domain.

Lastly, "Creating a Large Language Model of a Philosopher" investigates the potential of GTP in generating philosophical texts by fine-tuning GPT-3 with philosopher Daniel C. Dennett's works. The results show that both experts on Dennett's work and ordinary research participants struggle to distinguish machine-generated answers from those of the human philosopher. For two of the ten questions, the language model produced at least one answer that experts selected more frequently than Dennett's own answer. This fascinating outcome warrants further research to enhance the authenticity of machine-generated philosophy and explore other potential applications of GTP in the humanities.

In conclusion, the research articles discussed in this review provide significant insights into various aspects of ChatGPT. These studies collectively emphasize the potential benefits and applications of GPT in diverse areas such as predicting public opinion, grammatical error correction, translation, and even philosophy. However, they also uncover limitations and challenges that need to be addressed, including over-correction issues in GEC, limitations in translating low-resource or distant languages, and potential biases in predicting public opinion.

One common thread among these research articles is the need for further investigation and improvement of GTP to optimize their performance and minimize potential biases. As GPT becomes increasingly sophisticated and integrated into various industries, it is essential to consider the ethical implications of its use, including concerns about job displacement, fairness, and transparency.

Future research should focus on exploring innovative applications of GPT, addressing their limitations, and developing strategies to mitigate potential negative consequences. Additionally, interdisciplinary collaboration between researchers, practitioners, and policymakers can lead to a more comprehensive understanding of GPT and its implications in various domains. This will ensure the responsible development and deployment of GPT, maximizing its benefits while minimizing risks. As the field of AI and language modeling continues to evolve rapidly, staying informed about the latest research findings is crucial for understanding the broader implications of these technologies. This review has highlighted some of the most significant results (according to Altmetric) and findings in the current research landscape, emphasizing both the potential and the challenges associated with ChatGPT.

# Conclusion

In conclusion, GPT has undoubtedly revolutionized the realm of information dissemination and has itself become a focal point of scientific inquiry. As witnessed during the COVID-19 pandemic, traditional academic publishing approaches are progressively losing ground to more agile methods. Platforms such as preprints, Twitter, and Reddit have emerged as effective conduits for knowledge distribution in this rapidly evolving domain. To stay updated on the latest advancements in GPT, it is recommended to utilize these alternative channels for information acquisition (Petiška, 2023)."

1) Twitter search for GPT[3] (many tweets are relevant to scientific audiences as a new tool for scientists – Perplexity[4]);
2) Reddit search for GPT[5] and the forum ChatGPT[6], where you can find weekly updates[7];
3) Google News (e.g., US edition) for ChatGPT[8];
4) Google Scholar (allintitle:ChatGPT)[9]
5) Dimensions (can be filtered according to Altmetric and scientific areas to identify the most discussed research articles in your field)[10]

*We would like to acknowledge the assistance of an AI language model, ChatGPT in conducting this review article.*

---

[3] https://twitter.com/search?q=GPT&src=typed_query
[4] https://twitter.com/Artifexx/status/1645303838595858432
[5] https://www.reddit.com/search/?q=GPT
[6] https://www.reddit.com/r/ChatGPT/
[7] https://www.reddit.com/r/ChatGPT/comments/12diapw/gpt4_week_3_chatbots_are_yesterdays_news_ai/
[8] https://news.google.com/search?q=chatGPT&hl=en-US&gl=US&ceid=US%3Aen
[9] https://scholar.google.com/scholar?hl=en&as_sdt=0%2C5&q=allintitle%3AChatGPT&btnG=
[10] https://app.dimensions.ai/discover/publication?search_mode=content&search_text=chatgpt&search_type=kws&search_field=text_search&order=altmetric&and_facet_for=80003

# References

- Aydın, Ö., & Karaarslan, E. (2022). OpenAI ChatGPT generated literature review: Digital twin in healthcare. Available at SSRN 4308687.
- Booth, A. (2016). Searching for qualitative research for inclusion in systematic reviews: a structured methodological review. Systematic Reviews, 5(1), 74.
- Borenstein, M., Hedges, L. V., Higgins, J. P., & Rothstein, H. R. (2009). Introduction to meta-analysis. John Wiley & Sons.
- Cook, D. J., Mulrow, C. D., & Haynes, R. B. (1997). Systematic reviews: synthesis of best evidence for clinical decisions. Annals of Internal Medicine, 126(5), 376-380.
- Dwivedi, Y. K., Kshetri, N., Hughes, L., Slade, E. L., Jeyaraj, A., Kar, A. K., ... & Wright, R. (2023). "So what if ChatGPT wrote it?" Multidisciplinary perspectives on opportunities, challenges and implications of generative conversational AI for research, practice and policy. International Journal of Information Management, 71, 102642.
- George, A. S., & George, A. H. (2023). A review of ChatGPT AI's impact on several business sectors. Partners Universal International Innovation Journal, 1(1), 9-23.
- Golder, S., Loke, Y., & McIntosh, H. M. (2008). Room for improvement? A survey of the methods used in systematic reviews of adverse effects. BMC Medical Research, 8(1), 111.
- Grant, M. J., & Booth, A. (2009). A typology of reviews: An analysis of 14 review types and associated methodologies. Health Information & Libraries Journal, 26(2), 91-108.
- Green, B. N., Johnson, C. D., & Adams, A. (2006). Writing narrative literature reviews for peer-reviewed journals: secrets of the trade. Journal of Chiropractic Medicine, 5(3), 101-117.
- Gough, D., Oliver, S., & Thomas, J. (2017). An introduction to systematic reviews. Sage Publications.
- Higgins, J. P., Altman, D. G., Gøtzsche, P. C., Jüni, P., Moher, D., Oxman, A. D., ... & Sterne, J. A. (2011). The Cochrane Collaboration's tool for assessing risk of bias in randomised trials. BMJ, 343, d5928.
- Moher, D., Liberati, A., Tetzlaff, J., Altman, D. G., & The PRISMA Group. (2009). Preferred reporting items for systematic reviews and meta-analyses: The PRISMA statement. PLoS Medicine, 6(7), e1000097.
- Petiška, E. (2023). ChatGPT cites the most-cited articles and journals, relying solely on Google Scholar's citation counts.  As a result, AI may amplify the Matthew Effect in environmental science. Arxiv. https://arxiv.org/abs/2304.06794
- Sallam, M. (2023, March). ChatGPT Utility in Health Care Education, Research, and Practice: Systematic Review on the Promising Perspectives and Valid Concerns. In Healthcare (Vol. 11, No. 6, p. 887). MDPI.
- Thompson, N., & Hanley, D. (2018). Science is shaped by Wikipedia: evidence from a randomized control trial.
- Wang, S., Scells, H., Koopman, B., & Zuccon, G. (2023). Can chatgpt write a good boolean query for systematic review literature search?. arXiv preprint arXiv:2302.03495.
- Wells, G. A., Shea, B., O'Connell, D., Peterson, J., Welch, V., Losos, M., & Tugwell, P. (2000). The Newcastle-Ottawa Scale (NOS) for assessing the quality of nonrandomised studies in meta-analyses. Retrieved from http://www.ohri.ca/programs/clinical_epidemiology/oxford.asp